\documentclass{aa} 
\newcommand{\hrieuv}{\text{HRI}_{\text{EUV}}}
\usepackage{graphicx}
\usepackage{bm}
\usepackage{natbib}
\usepackage{mathtools}
\usepackage{amsmath}
\usepackage{multirow}
\usepackage[export]{adjustbox}
\usepackage{rotating}
\usepackage{tikz}
\usepackage{txfonts}
\usepackage{textcomp}
\usepackage{color}

\begin{document} 

   \title{Quasi-periodic pulsations in extreme-ultraviolet brightenings}
   

   \author{%
            {Daye Lim,}\inst{\ref{aff:CmPA}, \ref{aff:ROB}}\orcid{0000-0001-9914-9080}
            {Tom Van Doorsselaere,}\inst{\ref{aff:CmPA}}\orcid{0000-0001-9628-4113}
            {David Berghmans}\inst{\ref{aff:ROB}}\orcid{0000-0003-4052-9462}
            {Laura A. Hayes}\inst{\ref{aff:DIAS}}\orcid{0000-0002-6835-2390}
            {Cis Verbeeck}\inst{\ref{aff:ROB}}\orcid{0000-0002-5022-4534}
            {Nancy Narang}\inst{\ref{aff:ROB}}\orcid{0000-0001-9398-2063}
            {Marie Dominique}\inst{\ref{aff:ROB}}\orcid{0000-0001-9398-2063}
            \and
            {Andrew R. Inglis}\inst{\ref{aff:gsfc}, \ref{aff:catholic}}\orcid{0000-0003-0656-2437}
          }
 
   \institute{%
             \label{aff:CmPA}{Centre for mathematical Plasma Astrophysics, Department of Mathematics, KU Leuven, Celestijnenlaan 200B, 3001 Leuven, Belgium} \email{daye.lim@kuleuven.be}
            \and
            \label{aff:ROB}{Solar-Terrestrial Centre of Excellence – SIDC, Royal Observatory of Belgium, Ringlaan -3- Av. Circulaire, 1180 Brussels, Belgium}
            \and
            \label{aff:DIAS}{Astronomy \& Astrophysics Section, School of Cosmic Physics, Dublin Institute for Advanced Studies, DIAS Dunsink Observatory, Dublin, D15 XR2R, Ireland.}
            \and
            \label{aff:gsfc}{Solar Physics Laboratory, Heliophysics Science Division, NASA Goddard Space Flight Center, Greenbelt, MD 20771, USA}
            \and
            \label{aff:catholic}{Physics Department, The Catholic University of America, Washington, DC 20064, USA}
             }

   \authorrunning{Lim et al.}

  \abstract
   {Extreme-ultraviolet (EUV) observations have revealed small-scale transient brightenings that may share common physical mechanisms with larger-scale solar flares. A notable feature of solar and stellar flares is the presence of quasi-periodic pulsations (QPPs), which are considered a common and potentially intrinsic characteristic.}
   {We investigate the properties of QPPs detected in EUV brightenings, which are considered small-scale flares, and compare their statistical properties with those observed in solar and stellar flares.}
   {We extracted integrated light curves of 22,623 EUV brightenings in two quiet Sun regions observed by the Solar Orbiter/Extreme Ultraviolet Imager and identified QPPs in their light curves using Fourier analysis.}
   {Approximately 2.7\% of the EUV brightenings exhibited stationary QPPs. The QPP occurrence rate increased with the surface area, lifetime, and peak brightness of the EUV brightenings. The detected QPP periods ranged from approximately 15 to 260 seconds, which is comparable to the periods observed in solar and stellar flares. Consistent with observations of QPPs in solar and stellar flares, no correlation was found between the QPP period and peak brightness. However, unlike the trend observed in solar flares, no correlation was found between the QPP period and lifetime/length scale.}
   {The presence of QPPs in EUV brightenings supports the interpretation that these events may be small-scale manifestations of flares, and the absence of period scaling with loop length further suggests that standing waves may not be the primary driver of QPPs in these events.}


\maketitle

\section{Introduction}
Extreme-ultraviolet (EUV) observations have revealed transient, small-scale, localised brightness enhancements in the solar corona. These phenomena have been referred to by different names, such as blinkers \citep{1999A&A...351.1115H}, EUV bursts \citep{2021A&A...647A.159C}, EUV brightenings \citep{1998A&A...336.1039B}, nanoflares \citep{2000ApJ...529..554P, 2002ApJ...572.1048A, 2022A&A...661A.149P}, and campfires \citep{2021A&A...656L...4B, 2023A&A...671A..64D, narang25}, reflecting variations in the instruments and methods used for their detection. \citet{2003A&A...409..755H} demonstrated that these terms describe the same type of events. In this study, we hereafter collectively refer to them as EUV brightenings.

Imaging instruments with increasingly higher spatial resolution have enabled the detection of progressively smaller EUV brightenings, with the High Resolution Imager at 174 \AA\ of the Extreme Ultraviolet Imager (EUI $\hrieuv$; \citealt{2020A&A...642A...8R}) onboard Solar Orbiter recently resolving events as small as about 100 km \citep{narang25}. The temperature and emission measure (EM) of phenomena ranging from EUV brightenings to solar and stellar flares appear to follow a universal scaling law, implying that EUV brightenings may be small-scale flares that operate via similar physical mechanisms to those driving solar and stellar flares \citep{1999ApJ...526L..49S, 2002ApJ...577..422S, 2008ApJ...672..659A, 2023ApJ...943..143K, 2023MNRAS.522.4148K}. Thus, these EUV brightenings have attracted considerable interest due to their potential contribution to coronal heating \citep{1988ApJ...330..474P, 1991SoPh..133..357H}.

In addition, several pieces of evidence have suggested that EUV brightenings could be driven by magnetic reconnection. \citet{2021A&A...656L...7C} demonstrated that EUV brightenings seemed to be predominantly associated with plasma heating due to reconnection, based on a 3D radiation magnetohydrodynamic (MHD) simulation using the coronal extension of the MURaM code \citep{2005A&A...429..335V, 2017ApJ...834...10R}. \citet{2021ApJ...921L..20P} investigated the magnetic origin of 52 EUV brightenings detected with the Solar Orbiter/EUI $\hrieuv$, comparing them with line-of-sight magnetic field data from the Solar Dynamics Observatory/Helioseismic and Magnetic Imager (SDO/HMI; \citealt{2012SoPh..275..207S}). They found that most events resided on magnetic neutral lines where flux cancellation occurred. This result was further supported by similar characteristics observed in 38 brightenings detected by the Solar Orbiter/EUI $\hrieuv$ and the High Resolution Telescope of Polarimetric and Helioseismic Imager (PHI HRT; \citealt{2020A&A...642A..11S}) onboard Solar Orbiter, as reported by \citet{2022A&A...660A.143K}. 

A more recent study using a larger sample of EUV brightenings from Solar Orbiter/EUI $\hrieuv$ and SDO/HMI has yielded markedly different results. \citet{2024A&A...692A.236N} found that only about 12\% of the 5064 events occurred in bipolar regions. However, these findings could become clearer with future comparisons using high-resolution magnetograms. For instance, bipolar regions observed with the Swedish 1-m Solar Telescope/CRisp Imaging Spectropolarimeter (SST/CRISP; \citealt{2003SPIE.4853..341S, 2008ApJ...689L..69S}), which has a pixel scale of 0.057$^{\prime\prime}$, did not show the same characteristics in SDO/HMI (with a pixel scale of 0.5$^{\prime\prime}$), as discussed by \citet{2024A&A...686A.218N}. Additionally, an alternative mechanism that could explain the observed EUV brightenings has been proposed. \citet{2024ApJ...969L..34K} found through numerical simulations that EUV brightenings could be the coronal response to impulsive Alfvén waves in the chromosphere. Although the underlying physical mechanism of the observed EUV brightenings has not yet been clearly established, the possibility that they may share processes similar to those of standard flares remains noteworthy.

One of the key observational features of solar and stellar flares is the presence of quasi-periodic pulsations (QPPs). QPPs manifest as time-varying intensity variations in flare emissions that often have a quasi-periodic or periodic component, detectable across a broad spectrum of wavelengths, ranging from radio waves to gamma rays (see, e.g., \citealt{2016SoPh..291.3143V, 2021SSRv..217...66Z} for comprehensive reviews). In solar flares, periods of QPPs range from fractions of a second to several tens of minutes, while in stellar flares the QPP period can reach several hours \citep{2019ApJ...884..160V}. The governing mechanisms of QPPs depend on the type of QPPs observed but can be broadly categorised into three groups: MHD waves, repetitive magnetic reconnection (commonly referred to as magnetic dripping), and magnetic reconnection periodically triggered by external MHD waves \citep{2018SSRv..214...45M, 2020STP.....6a...3K}. Nevertheless, alternative interpretations remain plausible, necessitating further investigation.

Statistical studies on QPPs suggest that they are a common and potentially intrinsic feature of flares. \citet{2015SoPh..290.3625S} examined 35 X-class solar flares detected in soft X-ray (SXR) emissions and, using wavelet analysis, found that approximately 80\% of them exhibited QPPs. Similarly, \citet{2018SoPh..293...61D} applied wavelet techniques to 90 solar flares stronger than M5-class, observed in EUV and SXR emissions, and reported that around 90\% showed QPPs. \citet{2020ApJ...895...50H} investigated QPP occurrence rates in nearly 5600 solar flares observed in Geostationary Operational Environmental Satellite (GOES) SXR emissions using Fourier analysis and found that 9\% of C-class, 29\% of M-class, and 46\% of X-class flares displayed QPPs. Additionally, approximately 50\% of these M- and X-class flares exhibited non-stationary QPPs \citep{2023MNRAS.523.3689M}. In contrast, solar flares observed in radio \citep{2012ApJ...748..140M} and hard X-ray \citep{2016ApJ...833..284I} emissions showed occurrence rates of QPPs below 10\%. Stellar flares had QPP occurrence rates of less than 5\% \citep{2015MNRAS.450..956B, 2016MNRAS.459.3659P}. The discrepancy in occurrence rates stems from different criteria for identifying QPPs (i.e. stationary periodic or non-stationary), differences in instrument sensitivity, and/or wavelength bands. However, the finding that QPPs are not rare events remains significant, highlighting the need for further research.

Given the consistent scaling law observed from EUV brightenings to stellar flares \citep{2023ApJ...943..143K, 2023MNRAS.522.4148K}, as well as the frequent occurrence of QPPs in solar and stellar flares, it is reasonable to hypothesise that QPPs could also manifest in EUV brightenings. QPPs have already been observed in weak flare events, such as microflares \citep{2018ApJ...859..154N}, and intensity oscillations have been detected in coronal bright points in X-rays and EUV \citep{2008A&A...489..741T,  2013SoPh..286..125C, 2015ApJ...806..172S, 2023A&A...679A..10R}, which are larger than EUV brightenings and are also known to be associated with magnetic reconnection \citep{2011A&A...529A..21K, 2011A&A...526A.134A, 2015ApJ...810..163C, 2023ApJ...958L..38N}. In this paper, we aim to investigate whether QPPs are present in EUV brightenings, observed at the highest available resolution through the Solar Orbiter/EUI $\hrieuv$, which is considered to represent the smallest scale of flares. Additionally, we seek to compare the statistical properties of QPPs observed in small-scale flares with those found in solar and stellar flares.

\section{Data and Analysis}\label{sec:data}

\begin{figure*}[t]
  \resizebox{\hsize}{!}{\includegraphics{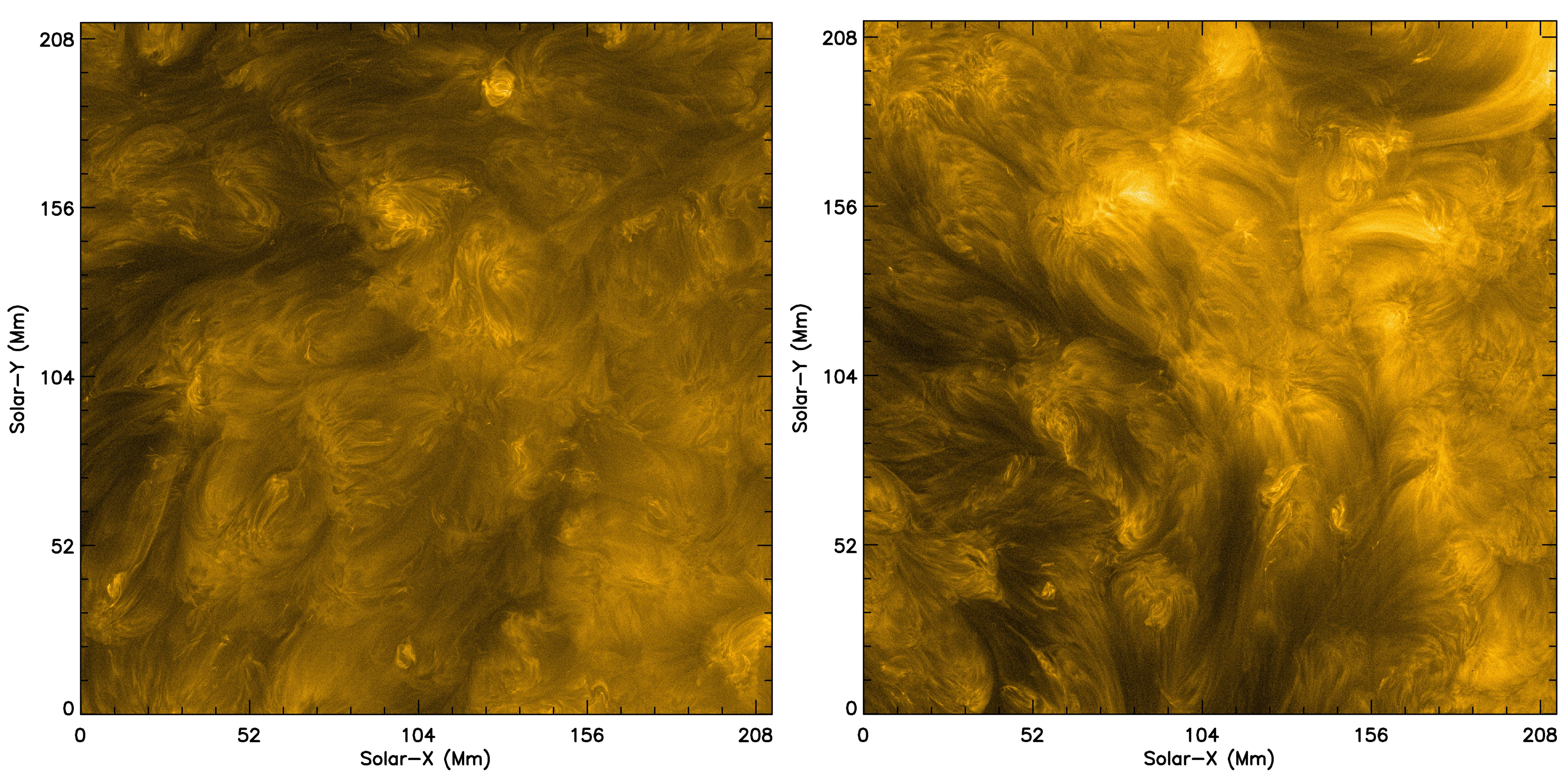}}
  \caption{Full FOV images from the Solar Orbiter/EUI $\hrieuv$ observed on 2022-10-12T05:25:00 (left) and 2023-10-06T10:00:00 (right). Both images have been processed using Multiscale Gaussian Normalization \citep{2014SoPh..289.2945M} to enhance their features.}
  \label{fig:hri}
\end{figure*}

\begin{figure*}[t]
  \resizebox{\hsize}{!}{\includegraphics{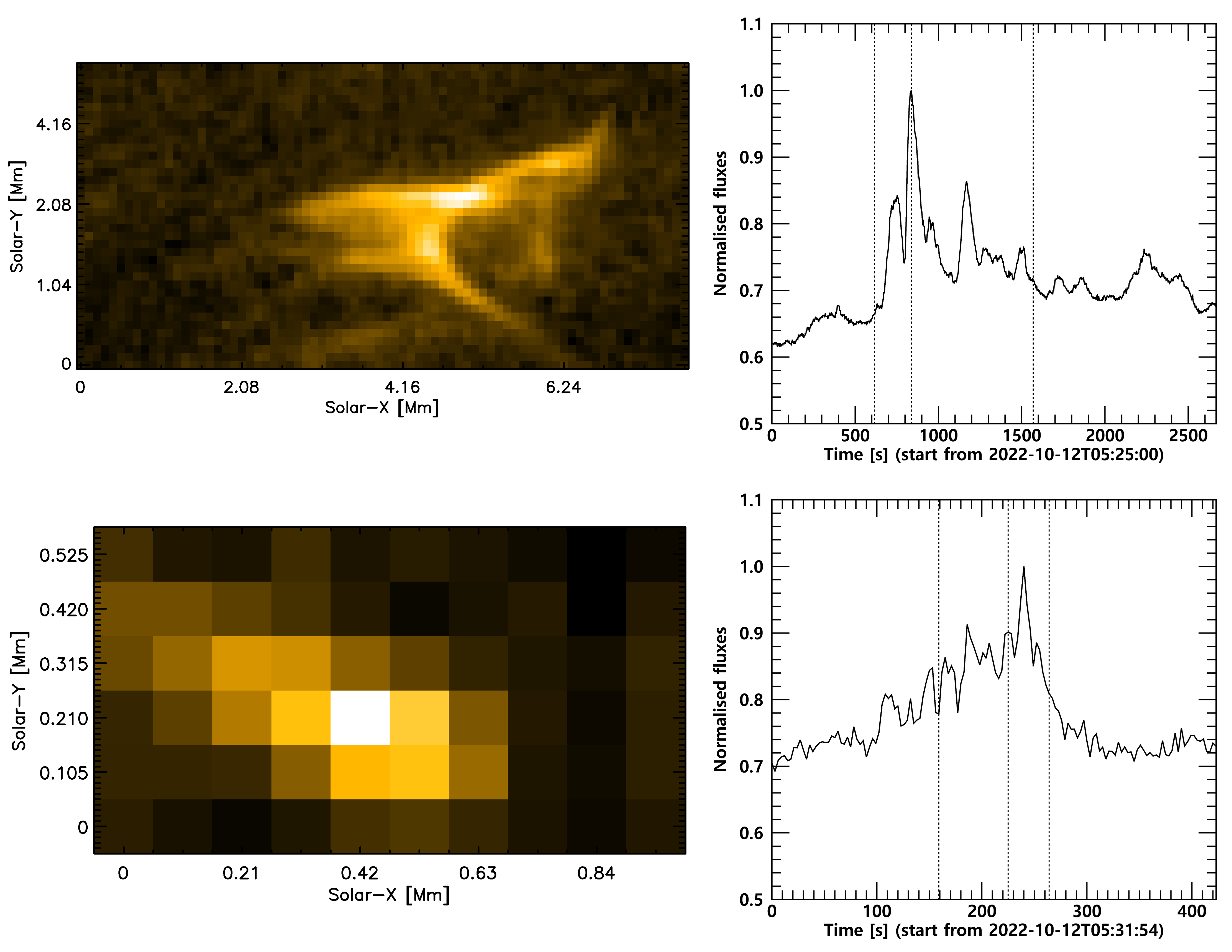}}
  \caption{Examples of EUV brightenings detected with $\hrieuv$ (left) alongside their corresponding light curves (right) showing the normalised integrated brightness over the entire brightening regions observed on 2022-10-12. Dashed lines indicate the start, peak, and end times of each EUV brightening. Each brightening image corresponds to its peak time.
  (An animation of this figure is available.)}
  \label{fig:intensity}
\end{figure*}
\subsection{Data}

Among the calibrated level-2 Solar Orbiter/EUI $\hrieuv$ data \citep{euidatarelease6}, we selected two quiet Sun regions observed at perihelion (approximately 0.29 au from the Sun) on 12 October 2022 and 6 October 2023, where the pixel plate scale was approximately 105 km. Both datasets were acquired with a cadence of 3 s. The observations on 12 October 2022 lasted for about 45 minutes, commencing at 05:25:00 UTC, while those on 6 October 2023 spanned 1 hour, starting at 10:00:00 UTC (see Figure \ref{fig:hri}). 

Efforts to detect small-scale EUV brightenings in high-resolution $\hrieuv$ observations have been ongoing, with the same automated algorithm consistently employed to ensure compatibility with previous studies \citep{2023A&A...671A..64D, 2024A&A...688A..77D, 2023A&A...676A..64N,  2024A&A...692A.236N}. This algorithm, originally developed by \citet{2021A&A...656L...4B} for detecting EUV brightenings in $\hrieuv$ data, defines the detected events by significant coefficients at the first two scales of an à trous wavelet decomposition, using a B3 spline scaling function. The amplitude at each scale is considered significant when they are greater than 6 times the root-mean-square amplitude expected from the noise model, considering photon shot noise and detector read noise (for further details, see \citealt{2021A&A...656L...4B, narang25}). Recently, \citet{narang25} applied this automated algorithm to the two datasets considered in this study, identifying a total of 177,738 events, including the smallest sample of EUV brightenings detected to date. In this study, we used the event catalogue from \citet{narang25}, selecting 22,623 brightenings with lifetimes exceeding 5 time frames (the strict limit based on the Nyquist criteria) for QPP analysis, thereby ensuring consistency and continuity with previous studies. To account for the smallest-scale brightenings, no selection criteria were imposed on the area, meaning that the smallest and shortest-lived events included in our dataset can be of one-pixel brightenings lasting 5 time frames. The lifetime of the EUV brightenings considered ranges from 6 to 890 frames.

The detected EUV brightenings exhibit variations in the location and size of their bright regions over their lifetime (see the animation of Figure \ref{fig:intensity}). To detect QPP signatures, we considered normalised light curves generated by summing the brightness in the region where the EUV brightening's maximum size was observed. To ensure that the light curve fully captures the temporal evolution of the brightening, we integrated the brightness over the time interval between $t_{\text{c}}-\tau$ and $t_{\text{c}}+\tau$, where $t_{\text{c}}$ and $\tau$ correspond to the event's central time and lifetime, respectively, as provided in the EUV brightening catalogue \citep{narang25}. Examples of the integrated light curves of the EUV brightenings are shown in the right panels of Figure \ref{fig:intensity}. Among the EUV brightenings analysed, the light curves of the larger event (approximately 8 Mm in the $x$-direction) and the smaller event (approximately 1 Mm in the $x$-direction) show distinctly different characteristics. For the larger event, the rising phase (the time interval from the onset of flux increase to the peak brightness) is relatively more rapid compared to the decay phase (the time interval from the peak brightness to the point where the flux returns to near its initial value). In contrast, the smaller event displays a more gradual rising phase and a relatively rapid decay phase. In both cases, the light curves do appear similar to solar flare light curves, typically observed in the impulsive phase of flares. \citet{2022Univ....8..471M} conducted a statistical analysis of the time profiles of the Reuven Ramaty High-Energy Solar Spectroscopic Imager \citep{2002SoPh..210....3L} SXR flares during Solar Cycles 23 and 24, categorising them into two groups: sudden flares, in which the rising phase is shorter than the decay phase, and gradual flares, in which the decay phase is shorter than the rising phase. They found that sudden flares occur regardless of magnetic activity, while gradual flares are enhanced during low solar activity. Given the overall similarity in the time profiles of flare light curves observed in both SXR and EUV wavelengths \citep{2011SSRv..159...19F}, the distinct light curve trends observed in EUV brightenings may be related to magnetic activity. Additionally, the light curve of the smaller event closely resembles the light curve of larger solar flares observed in hard X-rays (see Figure 2 in \citealt{2024A&A...684A.215C}), suggesting that the EUV brightening light curve may represent only the impulsive phase of the event. This raises the possibility that different physical mechanisms govern smaller and larger EUV brightening events. By studying QPPs in small-scale events, we can establish a link between their characteristics and those of larger solar flares, helping to reveal underlying physical flaring processes across different scales. However, these findings are based on just two cases, and further investigation into global trends and their detailed discussion is beyond the scope of this study, though it will be addressed in future work.

\subsection{Analysis}
The lack of a strict definition for QPPs, combined with their diverse characteristics and small amplitudes, complicates their detection. Some QPPs exhibit near-periodic behavior and can be analyzed with Fourier-based methods, while others are more irregular or time-varying, requiring alternative techniques. Consequently, a range of methods has been developed to search for QPP signatures.\citet{2019ApJS..244...44B} performed a blind test considering all such methods and found that the three approaches, Automated Flare Inference of Oscillations\footnote{\url{https://github.com/aringlis/afino_release_version}} (AFINO; \citealt{2015ApJ...798..108I, 2016ApJ...833..284I}), wavelet \citep{1998BAMS...79...61T}, and periodogram \citep{2017A&A...602A..47P}, yield robust results. However, \citet{2019ApJS..244...44B} pointed out that AFINO and periodogram tend to show a lower detection rate compared to other methods, suggesting that there might be more QPP events than detected by these methods. What this means, is that these Fourier based approaches are designed to identify stationary - more periodic - types of QPPs, and hence events that vary in period or phase are not detected. Nonetheless, AFINO offers unique advantages: it does not require preprocessing steps such as detrending, which could potentially lead to erroneous results in detected periods \citep{2016ApJ...825..110A, 2018SoPh..293...61D}, and it is particularly well-suited for handling large datasets effectively \citep{2020ApJ...895...50H} in a statistically robust manner. Therefore, we employed the AFINO code to investigate QPPs in EUV brightenings.
 
The AFINO method was developed to identify global QPP signatures within the flare time series. Its key feature lies in its analysis of the Fourier power spectrum of the flare signal, followed by a model fitting and comparison procedure to determine the most accurate representation of the data. A detailed description of AFINO can be found in \citet{2015ApJ...798..108I, 2016ApJ...833..284I} and \citet{2024ApJ...971...29I}. Here, we provide a brief overview of the essential information in the method.

The time series data of solar and stellar flares in the Fourier domain is commonly observed to follow a power law \citep{2011A&A...533A..61G}. In addition to this intrinsic characteristic, if there is a strong periodic component, it may manifest as a distinct peak in the Fourier spectrum. AFINO utilises these features to analyse the power spectral density of the flare light curve by considering and comparing several models. Before performing the analysis, the routine first normalises the light curve and applies a Hanning window to apodize the data. It then identifies the most appropriate model to determine whether a clear periodicity, indicative of QPPs, is present in the flare. The three models considered are: a power law ($S_{0}$), a power law with a Gaussian bump ($S_{1}$) which is to account for enhanced power at a specific frequency (i.e. signature of a quasi-periodic component in the lightcurve), and a broken power law ($S_{2}$). Among the three models, if a strong periodicity is present, Model $S_{1}$ is expected to provide the best fit. Conversely, if such periodicity is absent, either Model S0 or S2 would be more appropriate. AFINO employs the Bayesian Information Criterion (BIC), a statistical metric designed to assess the quality of a model by balancing its goodness of fit and complexity, to evaluate the model performance. A lower BIC value indicates a better fit relative to the other models being compared. Therefore, if the BIC of Model S1 is lower than those of Models S0 and S2, the EUV brightening can be considered a candidate exhibiting strong periodicity indicative of QPPs. Hence we calculate $\Delta$BIC = BIC$_{j}$ - BIC$_{S1}$, for $j=S0, S2$, and when $\Delta$BIC $>$ 0 we say that the QPP model is preferred.  

Since $\Delta$BIC indicates that Model $S_{1}$ provides the best relative fit to the power spectrum among the three models, we further assess its absolute suitability by considering a goodness-of-fit measure. The AFINO routine calculates the reduced chi-squared value for all models and provides the associated probability ($p$). Following the criterion suggested by \citet{2016ApJ...833..284I} and \citet{2020ApJ...895...50H}, we only accept cases when $p>0.01$. In summary, we classify a QPP as present when both $\Delta\text{BIC}_{S_{0}-S_{1}}$ and $\Delta\text{BIC}_{S_{2}-S_{1}}$ are greater than 0, and the $p$ for Model $S_{1}$ exceeds 0.01.

\section{Results and Discussion}\label{sec:results}

\begin{figure*}
  \resizebox{\hsize}{!}{\includegraphics{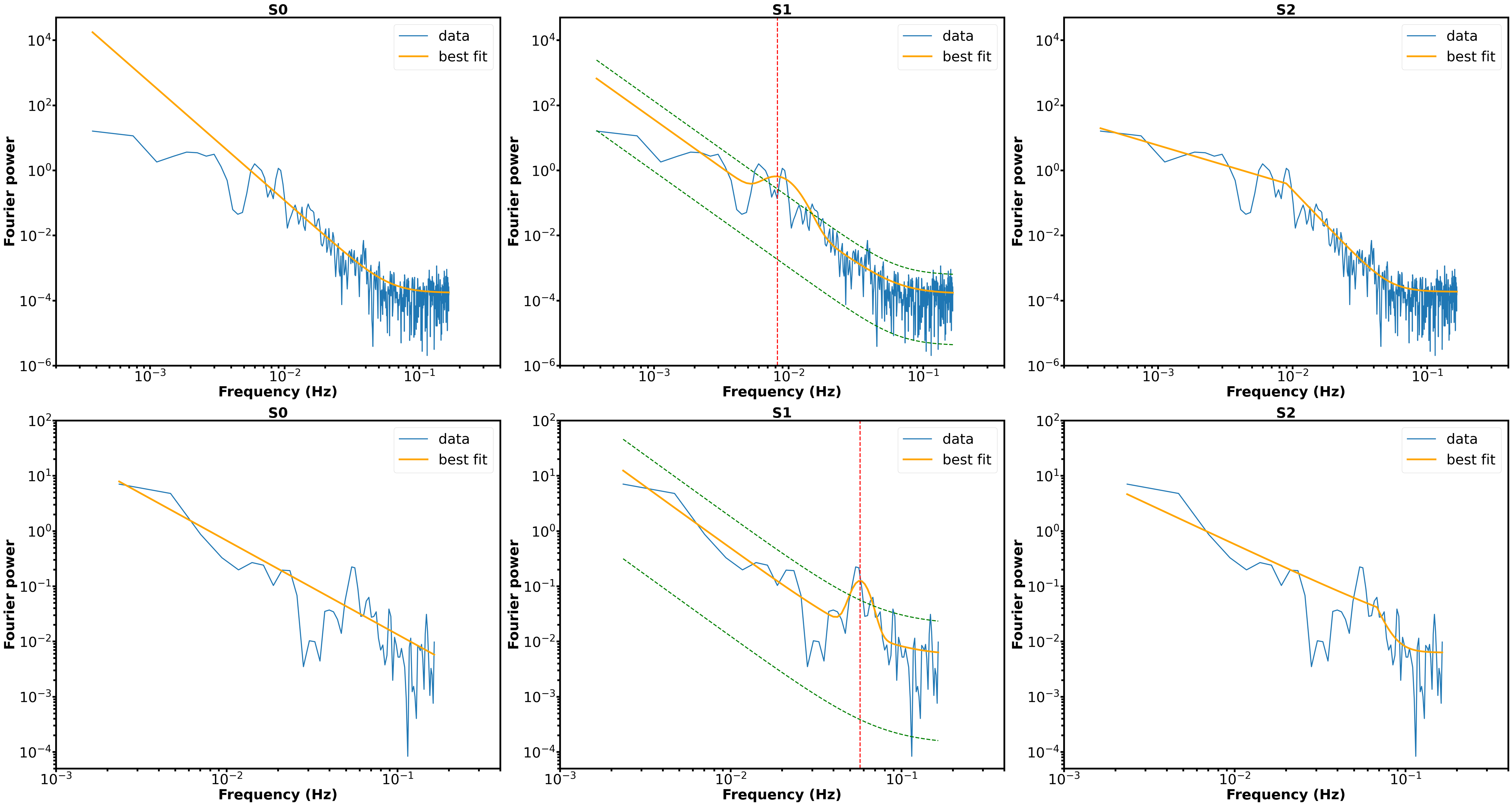}}
  \caption{Examples of AFINO results applied to the integrated light curves of the two EUV brightenings shown in Figure \ref{fig:intensity}, with the top and bottom panels corresponding to the respective brightenings in Figure \ref{fig:intensity}. The blue line represents the Fourier power of the light curve, while the orange line indicates the best fit for each model: $S_{0}$ (left), $S_{1}$ (centre) and $S_{2}$ (right). The red vertical dashed line in Model $S_{1}$ marks the Gaussian centre corresponding to the dominant period of about 121.17 s (top) and 17.43 s (bottom), and the green dashed lines represent the 95\% confidence interval. For the event in the top panel, the Bayesian Information Criterion (BIC) differences are $\Delta\text{BIC}_{S_{0}-S_{1}} = 16.6$ and $\Delta\text{BIC}_{S_{2}-S_{1}} = -36.8$. Model $S_{2}$ provides the best fit to the Fourier power, leading to its classification as a non-QPP event. In contrast, for the event in the bottom panel, Model $S_{2}$ is the most credible among the three models ($\Delta\text{BIC}_{S_{0}-S_{1}} = 11.3$ and $\Delta\text{BIC}_{S_{2}-S_{1}} = 10.1$). Since the probability for Model $S_{1}$ is 0.64, which exceeds the threshold of 0.01, this EUV brightening is classified as a QPP event.}
  \label{fig:afino}
\end{figure*}

After applying the AFINO code along with the two selection criteria for $\Delta$BIC and $p$ to the integrated light curves of 22,623 EUV brightenings, and considering only the detected periods longer than 15 seconds, we identified 602 cases exhibiting stationary QPP signatures. Figure \ref{fig:afino} presents examples of the AFINO results for the two EUV brightenings shown in Figure \ref{fig:intensity}. The Fourier analysis of the light curves for these events is represented by the blue curves, while the best fit of each model to power spectral density is shown in orange. For the larger event, the BIC differences were calculated as $\Delta\text{BIC}_{S_{0}-S_{1}} =16.6$ and $\Delta\text{BIC}_{S_{2}-S_{1}}=-36.8$. While the BIC for Model $S_{1}$ was lower than that for Model $S_{0}$, it remained higher than that for Model $S_{2}$. Therefore, this EUV brightening was not classified as exhibiting significant periodicity and was considered a non-QPP event. In contrast, for the smaller event, Model S1 provided the best fit among the three models, showing a high goodness of fit ($\Delta\text{BIC}_{S_{0}-S_{1}}=11.3$, $\Delta\text{BIC}_{S_{2}-S_{1}}=10.1$, and $p=0.64$). This EUV brightening was consequently classified as exhibiting QPP behaviour, with the detected periodicity of approximately 17.4 s.

\subsection{The occurrence rates of QPPs}

EUV brightenings, considered to be the smallest-scale flares observed to date, exhibited QPPs with constant periods in only approximately 2.7\% of events. This occurrence rate is substantially lower than the 90\% reported from Sun-as-a-star EUV observations \citep{2018SoPh..293...61D} and also lower than the average rate of 50–80\% observed in SXR emission \citep{2015SoPh..290.3625S, 2020ApJ...895...50H}. However, this low rate is comparable to the $\leq$5\% QPP occurrence rate observed in white-light emissions from stellar flares \citep{2015MNRAS.450..956B, 2016MNRAS.459.3659P}, which are typically larger and more intense than solar flares. It should be noted, however, that a recent study utilising high-cadence observations has reported the presence of QPPs in 30\% of white-light stellar flares \citep{2024arXiv241207580P}. While our result might suggest that strictly periodic QPPs are not prevalent in EUV brightenings, the derived occurrence rates could be influenced by observational biases or the inherent conservatism of the AFINO algorithm, which may lead to an underestimation. Therefore, a more definitive conclusion requires investigations of additional observational datasets and detection methods. Furthermore, across the scale from EUV brightenings to stellar flares, a comparison of these statistical results suggests a general lack of correlation between QPP occurrence rates and flare intensity or size. However, given the sensitivity of QPP detection to instrument sensitivity, temporal cadence, and detection methodology, statistically robust comparisons that mitigate these influences are necessary.

\begin{figure*}
  \resizebox{\hsize}{!}{\includegraphics{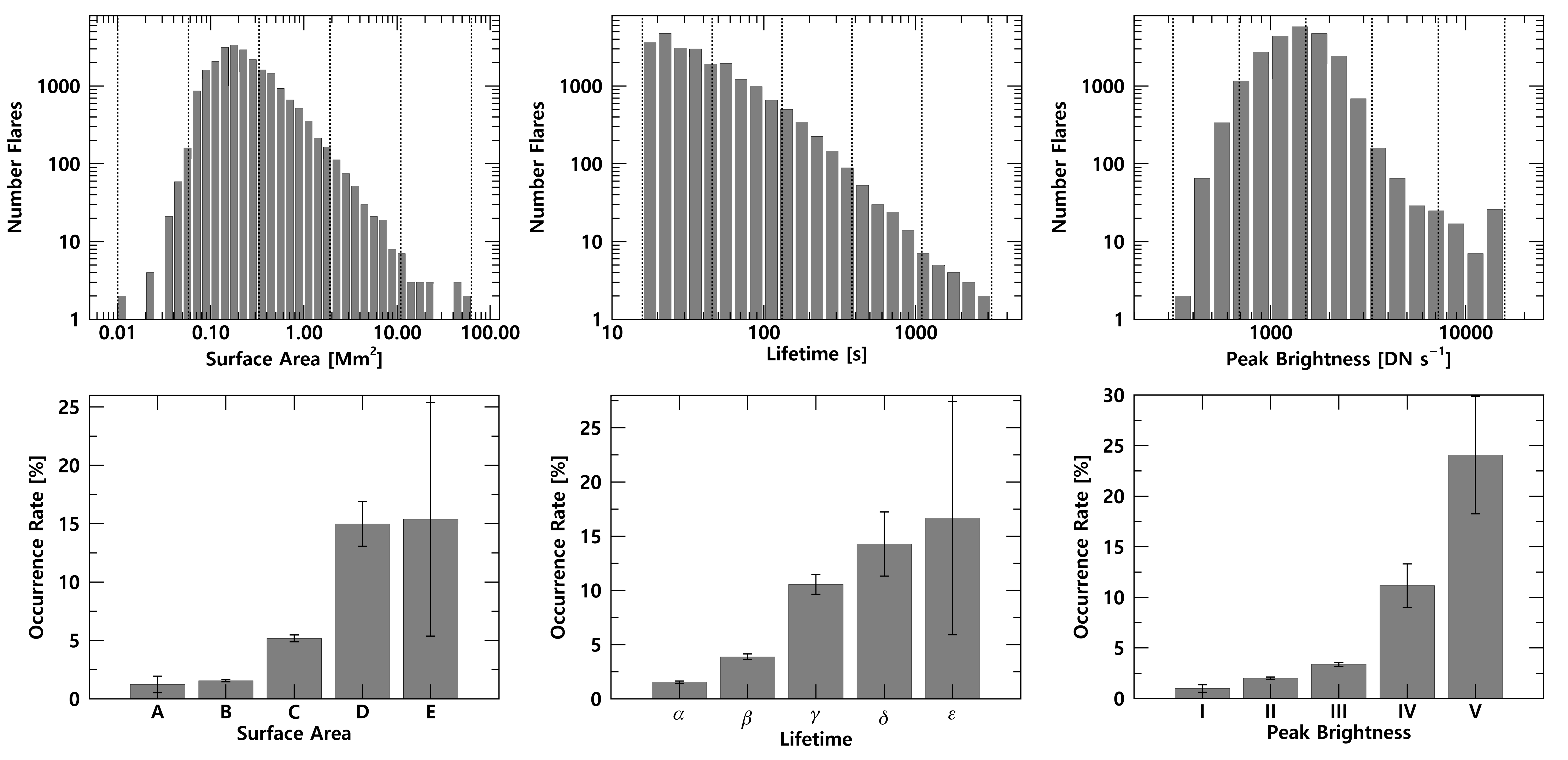}}
  \caption{Histograms of the surface area, lifetime, and peak brightness of EUV brightenings (top) and the occurrence rates of quasi-periodic pulsations (QPPs) as a function of these three parameters (bottom). A value of 1 was added as an offset to handle bins with a count of 1 in the logarithmic scale histogram. The dashed lines in the histograms indicate the threshold values used to divide the sample into five subgroups. The error bars for the occurrence rates, reflecting the varying bin populations, are calculated based on Poisson statistics.}
  \label{fig:rates}
\end{figure*}

Understanding the possible influence of EUV brightening characteristics on QPP occurrence requires a systematic approach. Therefore, we examine the occurrence rates of QPPs as a function of the surface area, lifetime, and peak brightness of EUV brightenings. To date, no established classification of EUV brightenings based on these three parameters exists. Therefore, to ascertain these rates, the dynamic range of each parameter, spanning its minimum and maximum observed values, was partitioned into five logarithmically equidistant bins (the top panels of Figure \ref{fig:rates}). These bins were designated as follows: for surface area, Group A (smallest) to Group E (largest); for lifetime, Group $\alpha$ to Group $\epsilon$; and for peak brightness, Group I to Group V. The QPP occurrence rate ($R$) for each bin was then determined by dividing the number of detected QPP events by the total number of EUV brightenings ($N$) within that respective bin, noting that each bin was not equally populated. Uncertainties ($\sigma=\sqrt{R(1-R)/N}$) associated with the occurrence rates, reflecting the varying bin populations, were estimated based on Poisson statistics. 

The observed occurrence rates ranged from $1.2\pm0.7$\% to $15.4\pm10.0$\% for the surface area, from $1.5\pm0.1$\% to $16.7\pm10.8$\% for the lifetime, and from $1.0\pm0.4$\% to $24.1\pm5.8$\% for peak brightness (the bottom panels of Figure \ref{fig:rates}). Furthermore, we considered the joint occurrence rate for the three parameters. The correlation coefficient (CC) between surface area and lifetime was approximately 0.8, indicating a strong positive correlation. However, peak brightness was found to be independent of both surface area and lifetime, with CCs of 0.35 in each case. No QPPs were detected for Group A$\alpha$I, while Group E$\epsilon$V exhibited a joint occurrence rate of $20.0\pm17.9$\%. For Group B$\alpha$III, the rate was $1.5\pm0.2$\%, whereas no QPP events were observed for Group C$\beta$I.
These results demonstrate a positive correlation between the QPP occurrence rate and the surface area, lifetime, and peak brightness of EUV brightenings, indicating a tendency for QPPs to occur more frequently in events that are larger, longer in lifetime, and of higher brightness. In standard solar flares, higher QPP occurrence rates were observed in groups with larger GOES X-ray peak fluxes [W $\text{m}^{-2}$] \citep{2020ApJ...895...50H}. While peak brightness and peak flux are not identical physical quantities, if we consider both parameters as approximate indicators of flare strength, the enhanced QPP occurrence rate observed in more powerful flares appears consistent across both large-scale solar flares and smaller-scale EUV brightenings. For C1.0-class flares, with an occurrence rate of approximately 4\% and estimated thermal energy ranging from approximately $10^{29}$ to $10^{30}$ erg \citep{2019ApJ...874..157R, 2023ApJ...958..104K}, extrapolation based on the observed trend in solar flares suggests that flares with energies below $10^{29}$ erg would exhibit occurrence rates lower than 4\%. In this study, the absence of simultaneous SDO/AIA observations for the EUV brightenings precluded the precise calculation of thermal energy. However, based on EM analysis performed by \citet{2021A&A...656L...4B}, these brightenings are estimated to possess energies between approximately $10^{21}$ and $10^{24}$ erg. While considering this energy range as a single group yields an occurrence rate of approximately 2.7\%, potentially suggesting a consistent trend, examining the occurrence rates within sub-groups based on the parameters influencing thermal energy reveals deviations from this trend. However, given the disparate sensitivities of the instruments involved and the considerable uncertainties associated with the estimated thermal energies, a quantitative and systematic analysis of occurrence rates spanning energies from the EUV brightening scale to the standard solar flare scale requires further investigation and should be revisited in future work. 

\subsection{The periods of QPPs}

\begin{figure*}
  \resizebox{\hsize}{!}{\includegraphics{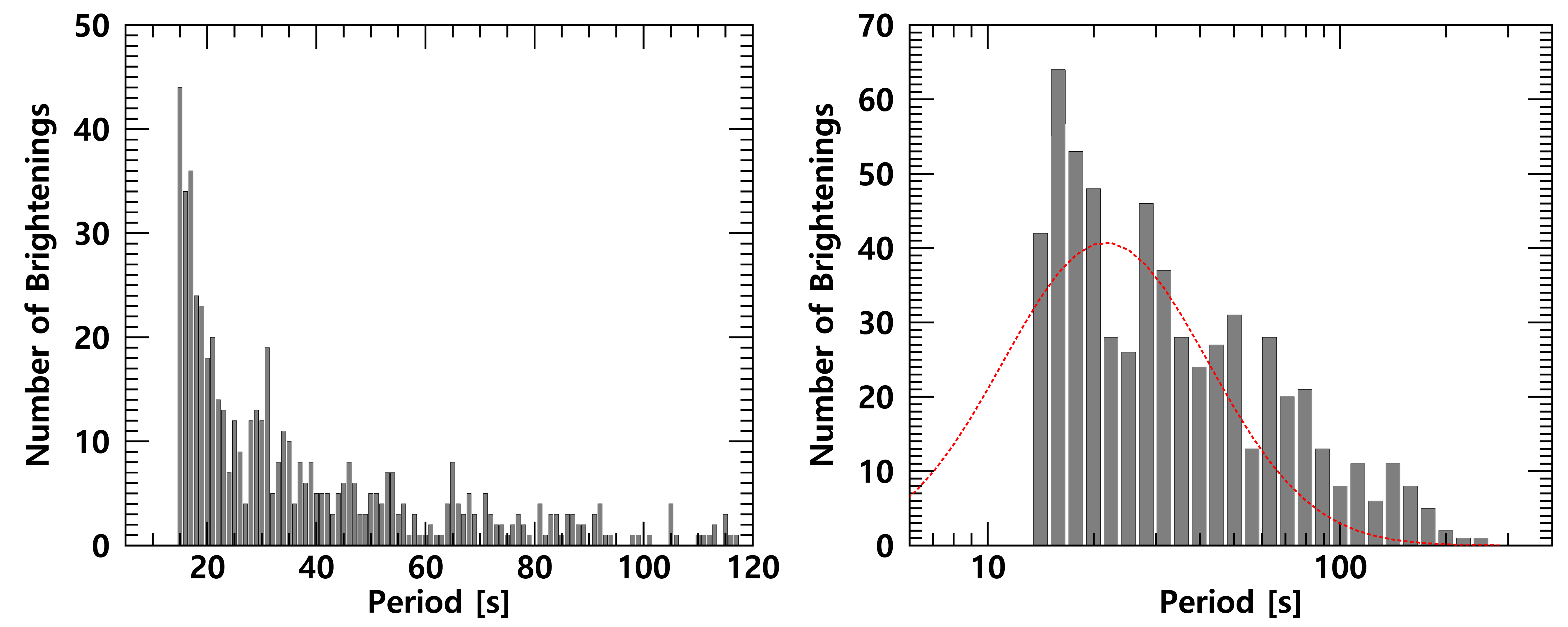}}
  \caption{Histograms of the periods of detected QPPs. Left panel: Histogram showing periods up to 120 s. Right panel: Logarithmic histogram showing the full range of periods. Note that periods below 15 s are excluded, as we applied a conservative criterion requiring detected periods to be at least five times the observational cadence of 3 s. The red dashed line represents the normal distribution in log space, with a mean of 21.6 s of QPP periods observed in standard solar flares \citep{2020ApJ...895...50H}.}
  \label{fig:hist_per}
\end{figure*}

The periods of QPPs detected in EUV brightenings range from 15 s to 255 s, with periodicities below 30 s appearing dominant (the left panel of Figure \ref{fig:hist_per}). QPP periods detected using wavelet analysis from EUV time series of 90 flares greater than M5-class ranged from approximately 1 s to 100 s, again with a dominance of periods shorter than 30 s \citep{2018SoPh..293...61D}. In addition, this period range exhibits agreement with the results obtained from 13 QPP events, the periods of which spanned from approximately 6 to 110 seconds, recently observed in white-light flares originating from M dwarfs \citep{2024arXiv241207580P}.
The observed similarity in QPP periods across these diverse flare scales (EUV brightenings, solar flares, and stellar flares) suggests that the QPP period is not strongly dependent on flare size or intensity. This will be discussed further in Section \ref{sec:scatter}. Furthermore, it is anticipated that QPPs with shorter periods, between 1 s and 15 s, could be detected in EUV brightenings observed with higher cadence data. QPP period statistics for a larger sample of flares observed in SXR exhibit somewhat different characteristics. QPP periods detected using AFINO by \citet{2020ApJ...895...50H} ranged from approximately 6 s to 300 s, with a lower cutoff at 6 s due to the observation cadence, and were well represented by a log-normal distribution with a mean of about 21 s and a 1$\sigma$ range between 10 s and 40 s on a logarithmic scale. This distribution differs from the distributions found in this study and by \citet{2018SoPh..293...61D}, which show a greater concentration of events near the detection limit and had an upper cutoff at 100 s imposed by the detection method. However, the long-period tail of this distribution shows good agreement with the period distribution observed in EUV brightenings (the right panel of Figure \ref{fig:hist_per}). As noted by \citet{2020ApJ...895...50H}, the presence of dominant QPP periodicities between 10 s and 60 s is observed regardless of the observing wavelength \citep{2018SSRv..214...45M}. This study further demonstrates that this characteristic is also independent of flare size. This suggests that the mechanism responsible for QPP generation may be independent of, or less sensitive to, flare size and/or temperature, highlighting the need for further investigation to clarify this.

\subsection{Statistical relationships between QPP periods and EUV brightening characteristics}\label{sec:scatter}

\begin{figure*}
  \resizebox{\hsize}{!}{\includegraphics{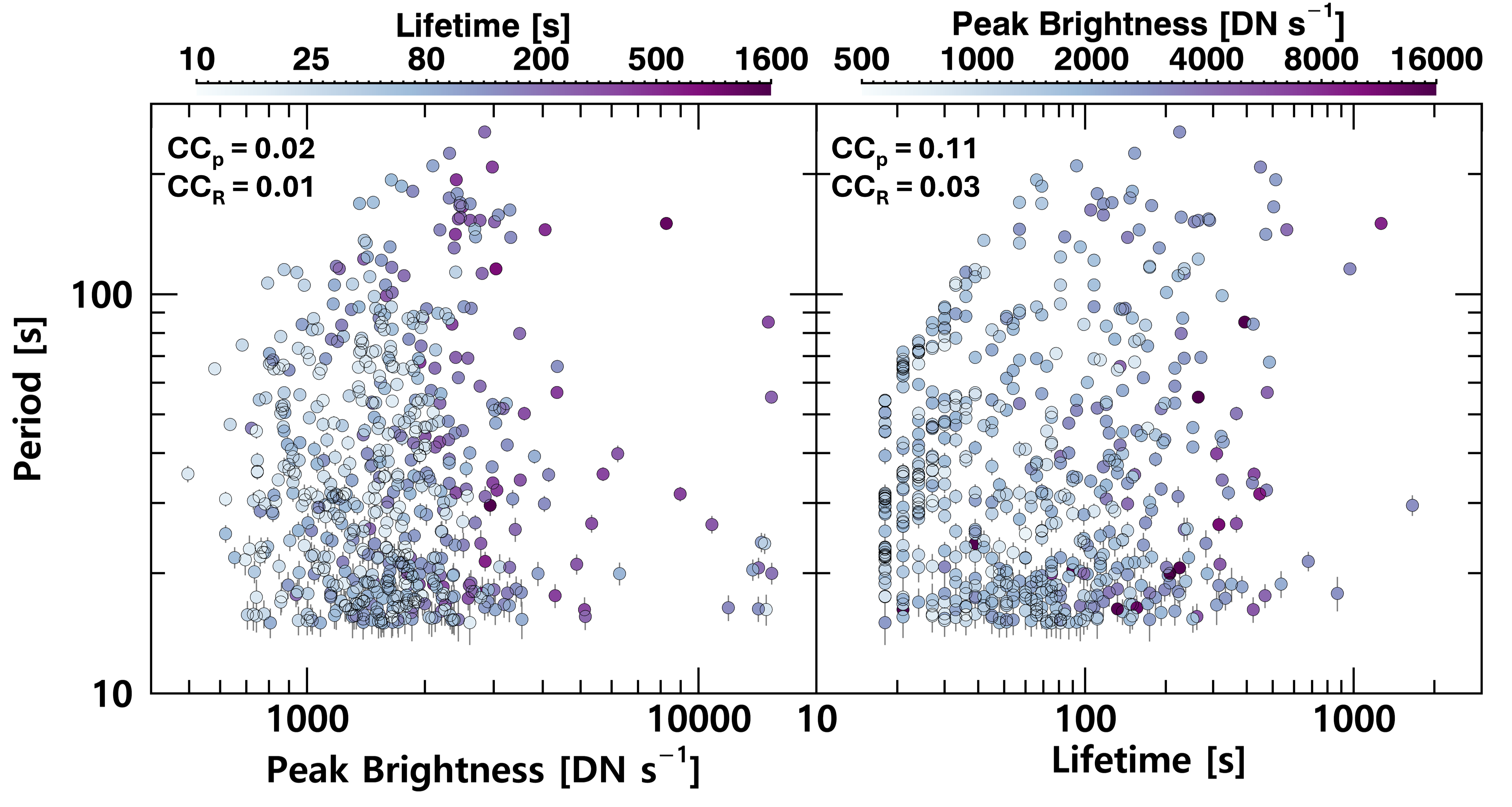}}
  \caption{Scatter plots of QPP period versus peak brightness (left panel) and lifetime (right panel) of EUV brightenings. Period uncertainties are estimated from the frequency width of the Gaussian bump in the model $S1$ to the power spectrum. The colour scales in each panel represent the logarithmic values of lifetime and peak brightness, respectively. The Pearson correlation coefficient ($\text{CC}_\text{P}$) and Spearman's rank correlation ($\text{CC}_\text{R}$) are indicated in the figure.}
  \label{fig:scatter_per_int_dur}
\end{figure*}

We investigate the relationship between QPP periods and the characteristics of EUV brightenings. Initially, we demonstrate a negligible correlation between period and peak brightness, with a Pearson correlation coefficient ($\text{CC}_{\text{P}}$) of 0.02 and a Spearman's rank correlation ($\text{CC}_{\text{R}}$) of 0.01 (the left panel of Figure \ref{fig:scatter_per_int_dur}). Furthermore, in the case of larger flares, the relationship between flare magnitude and QPP period was also found to be independent \citep{2016ApJ...833..284I, 2017A&A...608A.101P, 2018SoPh..293...61D, 2020ApJ...895...50H}. This independence extends to stellar QPPs, where statistical studies have shown no correlation between QPP periods and flare energy \citep{2016MNRAS.459.3659P, 2024arXiv241207580P}. While standing MHD waves could explain QPP periods independent of flare intensity, the similar period ranges observed from small EUV brightenings to large flares argue against this mechanism. Given that oscillatory reconnection, a potential QPP mechanism, exhibits periodicity independent of the amount of energy released \citep{2022ApJ...933..142K}, this mechanism may also be applicable to the QPPs observed in the EUV brightenings considered in this study.

\begin{figure*}
  \resizebox{\hsize}{!}{\includegraphics{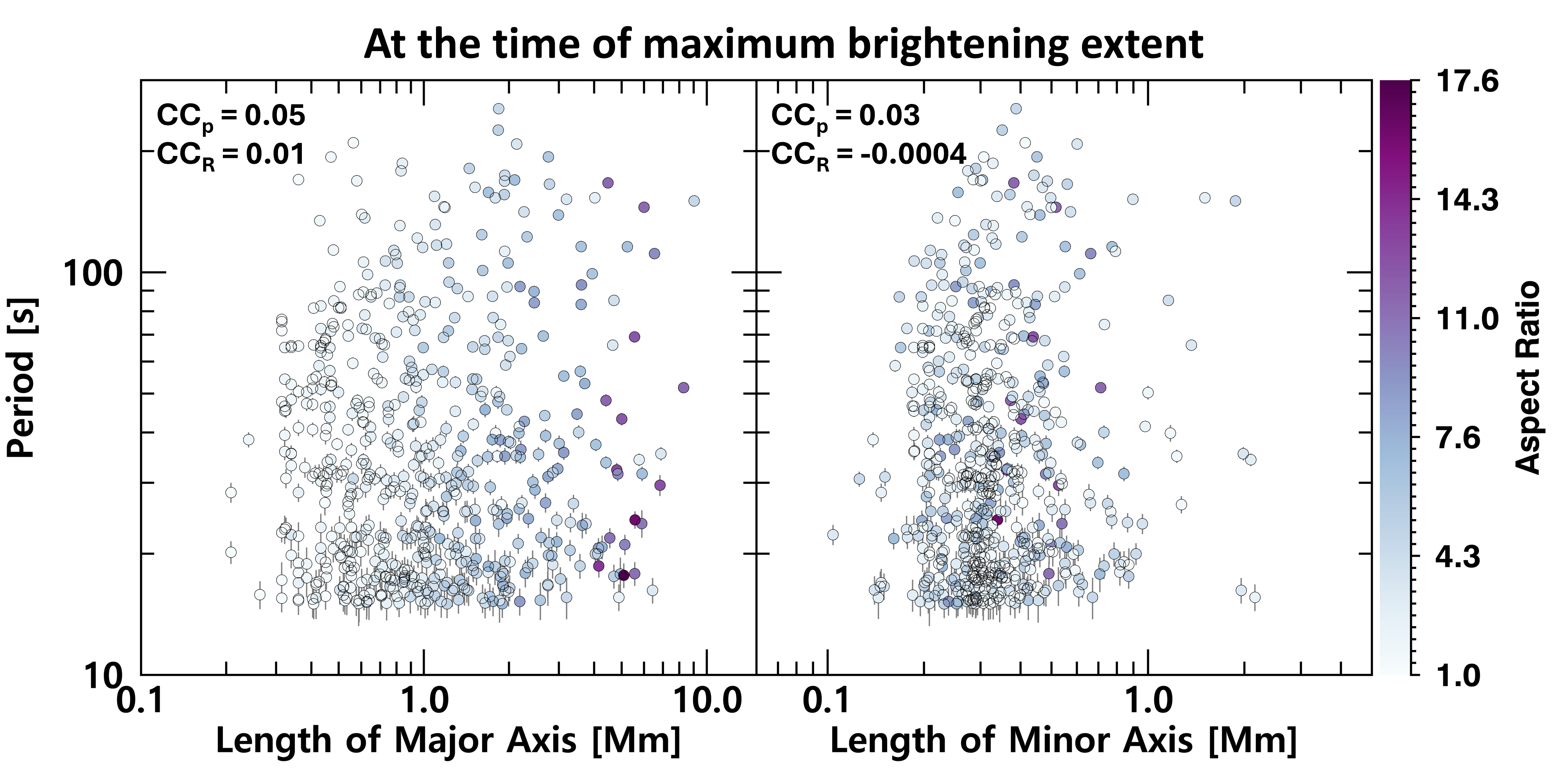}}
  \resizebox{\hsize}{!}{\includegraphics{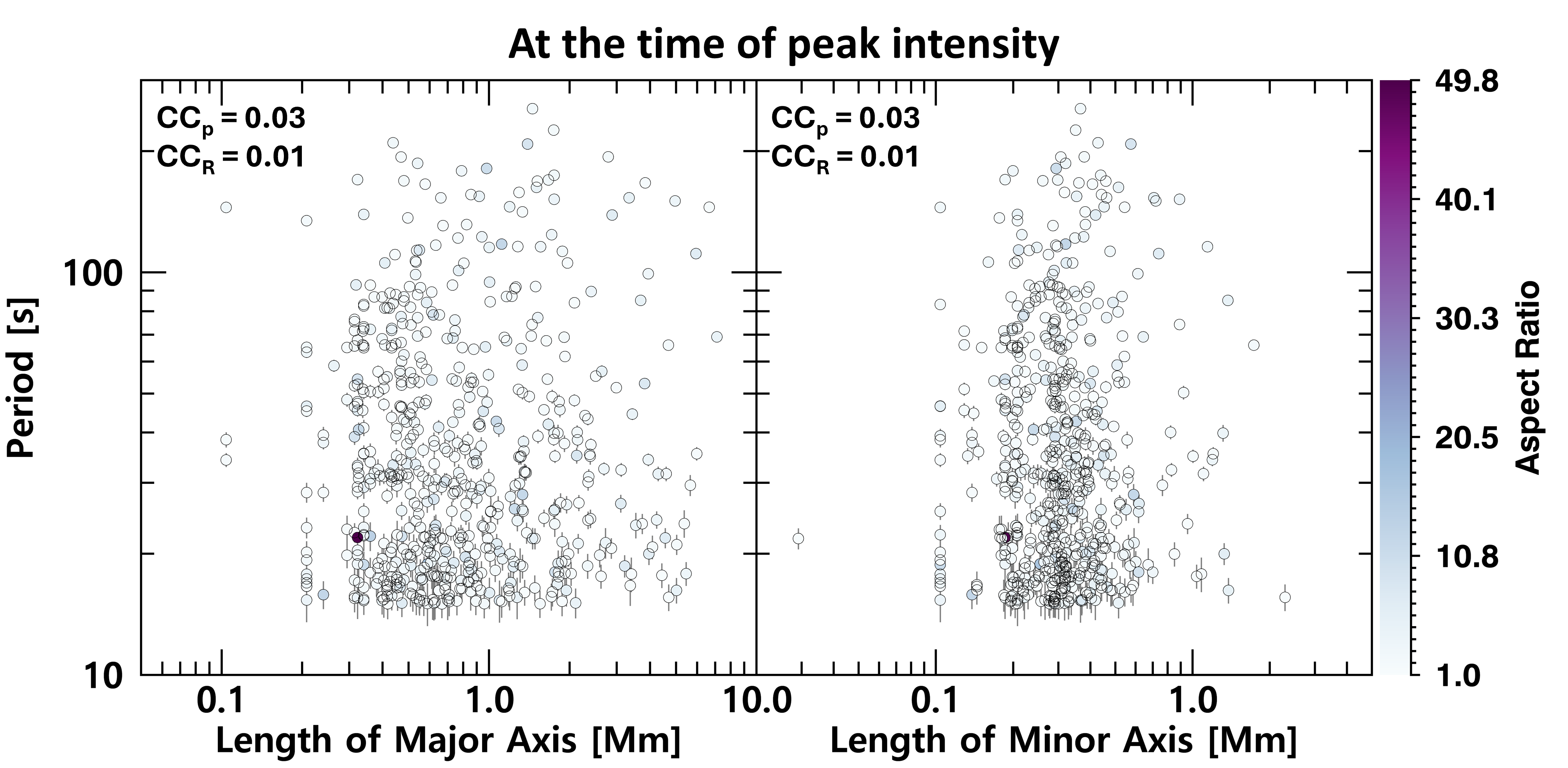}}
  \caption{Scatter plots of QPP period versus EUV brightening length scales. The top row compares QPP periods with the major (left panel) and minor (right panel) axis lengths at the time of maximum brightening extent. The bottom row shows the same comparison at the time of peak brightness. Period uncertainties are estimated from the frequency width of the Gaussian bump in the model $S1$ to the power spectrum. Colour coding indicates the aspect ratio (major axis length/minor axis length) in each case. The Pearson correlation coefficient ($\text{CC}_\text{P}$) and Spearman's rank correlation ($\text{CC}_\text{R}$) are indicated in the figure.}
  \label{fig:scatter_per_len}
\end{figure*}

The relationship between QPP period and lifetime, shown in the right panel of Figure \ref{fig:scatter_per_int_dur}, also reveals no significant correlation ($\text{CC}_{\text{R}}=0.03$ and $\text{CC}_{\text{P}}=0.11$). This contrasts with the strong correlation between flare duration and QPP period reported for GOES X-ray flares \citep{2019A&A...624A..65P, 2020ApJ...895...50H} and also for white light stellar flares \citep{2024arXiv241207580P}. The relationship between QPP period and the length scale of EUV brightenings (Figure \ref{fig:scatter_per_len}) also shows an opposite trend to that observed in larger solar flares \citep{2019A&A...624A..65P, 2020ApJ...895...50H}. Considering two representative length scales for the EUV brightening events, the maximum length attained during the event and the length scale at peak brightness, we found no significant correlation in either case ($\text{CC}_{\text{R}}$ and $\text{CC}_{\text{P}}$ both within $\pm0.05$). \citet{2020ApJ...895...50H} suggested that longer flare durations could lead to increased loop lengths due to successive magnetic reconnections, potentially influencing QPP periods. For EUV brightenings, we find a moderate correlation between lifetime and length (approximately 0.7 for maximum length and 0.6 for length at peak brightness), indicating that longer lifetimes are associated with longer loop lengths. These results suggest that standing waves, characterised by periods strongly influenced by loop length, are unlikely to be the primary mechanism responsible for QPPs observed in EUV brightenings.

However, given the rapid evolution of EUV brightenings, the two length scales considered above may not precisely reflect the scales directly associated with the QPP generation mechanism. To mitigate this ambiguity, we further investigate the relationship between QPP period and length scale. Considering the diverse morphologies of EUV brightenings, ranging from dot-like to loop-like structures \citep{2021A&A...656L...4B}, it is plausible that distinct generation mechanisms may be responsible for each type. 

\begin{figure*}
  \resizebox{\hsize}{!}{\includegraphics{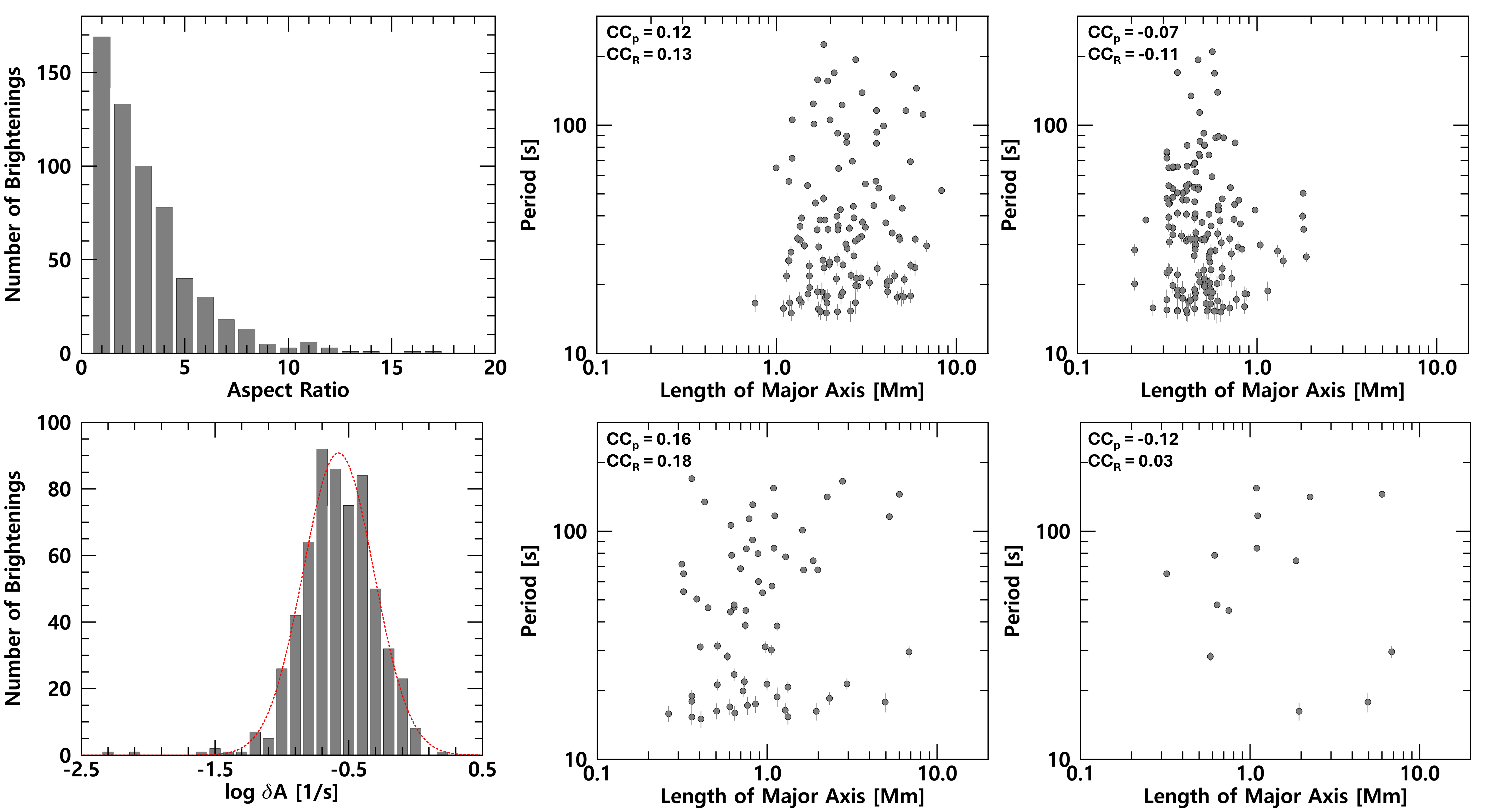}}
  \caption{Scatter plots of QPP period versus EUV brightening major axis length scales, depending on the aspect ratio (top) and the rate of area change, $\delta A$ (bottom). The left panels show histograms of the aspect ratio and log $\delta A$, respectively. In the histogram of log $\delta A$, the red dashed line indicates the fitted normal distribution with the mean ($\mu=-0.58$) and the standard deviation ($\sigma=0.27$). For the aspect ratio, scatter plots are shown for events with values greater than 5 (top centre) and less than 2 (top right). Similarly, for the rate of area change, scatter plots are presented for events with values smaller than $\mu-\sigma$ (bottom centre) and smaller than $\mu-2\sigma$ (bottom right). The Pearson correlation coefficient ($\text{CC}_\text{P}$) and Spearman's rank correlation ($\text{CC}_\text{R}$) are indicated in the figure.}
  \label{fig:scatter_per_len_add}
\end{figure*}

Consequently, exploring relationships based on morphological classification could provide a more accurate understanding. We therefore investigate the distribution of aspect ratios, calculated as the ratio of major to minor axis lengths, using the maximum length scale of each EUV brightening. The aspect ratios range from 1 to 18. Based on this distribution, we assumed that EUV brightenings with an aspect ratio greater than 5 exhibited a loop-like morphology, while those with an aspect ratio less than 2 were considered to be more akin to dot-like events. For each case, although the correlation was marginally improved compared to the overall sample, the relationship between the major axis length scale and period remains uncorrelated ($\text{CC}_{\text{R}}$ and $\text{CC}_{\text{P}}$ both within $\pm0.12$, see top panels of Figure \ref{fig:scatter_per_len_add}). 

During the lifetime of EUV brightening events, the area of smaller events remains relatively constant, whereas larger events can exhibit significant variations in area (see the animation of Figure \ref{fig:intensity}). As significant temporal variations in the area can complicate the precise definition of the length scale directly associated with QPPs, we restrict our analysis to a sample of events with relatively small rates of area change ($\delta A$ [1/s]),
\begin{equation}
    \delta A=\frac{\frac{A_{\text{max}}-A_{\text{min}}}{A_{\text{min}}}}{\tau},
\end{equation}
where $A_{\text{max}}$ and $A_{\text{min}}$ are the maximum area and minimum area during its lifetime, respectively.
From the histogram of log $\delta A$, we fitted a normal distribution and derived the mean ($\mu=-0.58$) and standard deviation ($\sigma=0.27$) of the distribution. Considering events with log $\delta A$ smaller than $\mu-\sigma$ resulted in a group of 65 events with a $\text{CC}_{\text{R}}$ of 0.18 and $\text{CC}_{\text{P}}$ of 0.16. Applying a stricter criterion ($\delta A$ smaller than $\mu-2\sigma$) yielded a smaller group of 14 events with a $\text{CC}_{\text{R}}$ of 0.03 and a $\text{CC}_{\text{P}}$ of -0.12. Although these CC values represent a slight increase compared to that obtained for the entire sample, they still indicate a negligible correlation (see bottom panels of Figure \ref{fig:scatter_per_len_add}).

Further analysis also reveals no discernible correlation between the length scale of EUV brightenings and the QPP period. As discussed above, while this result may be intrinsic to the mechanism responsible for the QPPs, we cannot exclude the possibility that the length scales considered in this study are not representative of the structures hosting these events. If the EUV brightening originates at the apex of a small loop \citep{2021A&A...656L...4B, 2021A&A...656L...7C}, our measurements would reflect the length of a portion of the apex, rather than the entire loop length. To obtain a more accurate estimation of the loop length, we could either estimate the precise height of each event using simultaneous SDO/AIA observations to infer a representative length \citep{2021A&A...656A..35Z}, or determine the footpoints of each event by comparing magnetogram and $\hrieuv$ data and then calculate the distance between them (assuming the loop hosting EUV brightening is semi-circular). 

\section{Conclusions}\label{sec:conclusion}

We analysed the integrated light curves of 22,623 EUV brightenings detected in two $\hrieuv$ datasets observing quiet Sun regions at perihelion. Among these, 602 events exhibiting a constant period were identified using Fourier analysis with the AFINO code. Stationary QPPs in EUV brightenings tended to occur more frequently in events with longer lifetimes, larger surface areas, and higher peak brightnesses (with an occurrence rate of approximately 20\% in the higher-value groups). The observed trend of QPP frequency increasing with brightness in EUV brightenings aligned with findings from studies of GOES X-ray solar flares. Notably, the detected QPP periods, ranging from 15 s to approximately 260 s, were comparable to those observed in both solar and stellar flares, despite the substantial differences in their respective scales and intensities. Furthermore, we investigated the relationship between the QPP period and the characteristics of the EUV brightenings. Consistent with results from standard solar flares, no correlation was found between the period and the peak brightness. In the case of EUV brightenings, no correlation was found between QPP periods and either event lifetime or length scale, a trend contrasting with that observed in solar flares. This implies that standing wave modes, typically associated with loop length-dependent periods, may not be the dominant mechanism underlying QPPs in EUV brightenings.

Our results demonstrate the robust presence of QPPs in EUV brightenings, suggesting that these events may be manifestations of flares. The comparable QPP occurrence rates and consistent statistical properties further support this interpretation.

A more comprehensive comparison of QPP characteristics across flares of varying scales and intensities requires larger data samples and further efforts to minimise the influence of instrument sensitivity and QPP detection methods. Furthermore, given the prevalence of non-stationary QPPs in solar and stellar flares \citep{2019PPCF...61a4024N, 2023MNRAS.523.3689M}, future work will include an analysis of non-stationary QPPs in EUV brightenings. While this study examined QPP characteristics over the entire timescale of the integrated light curves, future work will classify events by light curve morphology, separating those resembling typical SXR flares from other types. This will enable a deeper understanding of the mechanisms behind different EUV brightening types. One potentially relevant insight comes from studies of rapid pulsations in millimeter wavelengths during the impulsive phase of flares, which revealed a linear relationship between pulsation rate and flare flux, suggesting a link to quasi-periodic reconnection \citep{2009ApJ...697..420K, 2016SoPh..291.2003G}. The exploration of QPPs with periods shorter than the minimum (15 s) detected in this study will be possible with the release of Solar Orbiter/EUI 1-s cadence campaign data\footnote{\url{https://s2e2.cosmos.esa.int/confluence/display/SOSP/LTP17+Q4-2024}}, as well as through future campaigns. We may also anticipate this from future missions focused on high spatial and temporal resolution, such as the Multi-slit Solar Explorer (down to 0.5 s; \citealt{2022ApJ...926...52D}) and the Solar-C EUV High-Throughput Spectroscopic Telescope (up to 1 s; \citealt{2019SPIE11118E..07S}).

\begin{acknowledgements}
      Solar Orbiter is a space mission of international collaboration between ESA and NASA, operated by ESA. The EUI instrument was built by CSL, IAS, MPS, MSSL/UCL, PMOD/WRC, ROB, LCF/IO with funding from the Belgian Federal Science Policy Office (BELSPO/PRODEX PEA C4000134088, 4000112292 and 4000106864); the Centre National d’Etudes Spatiales (CNES); the UK Space Agency (UKSA); the Bundesministerium für Wirtschaft und Energie (BMWi) through the Deutsches Zentrum für Luft- und Raumfahrt (DLR); and the Swiss Space Office (SSO). The research that led to these results was subsidised by the Belgian Federal Science Policy Office through the contract B2/223/P1/CLOSE-UP. DL was supported by a Senior Research Project (G088021N) of the FWO Vlaanderen. TVD was supported by the C1 grant TRACEspace of Internal Funds KU Leuven and a Senior Research Project (G088021N) of the FWO Vlaanderen. Furthermore, TVD received financial support from the Flemish Government under the long-term structural Methusalem funding program, project SOUL: Stellar evolution in full glory, grant METH/24/012 at KU Leuven. The paper is also part of the DynaSun project and has thus received funding under the Horizon Europe programme of the European Union under grant agreement (no. 101131534). Views and opinions expressed are however those of the author(s) only and do not necessarily reflect those of the European Union and therefore the European Union cannot be held responsible for them. CV thanks the Belgian Federal Science Policy Office (BELSPO) for the provision of financial support in the framework of the PRODEX Programme of the European Space Agency (ESA) under contract numbers 4000143743 and 4000134088. MD acknowledges support from BELSPO in the framework of the ESA-PRODEX program, PEA 4000145189. LAH is funded through a Royal Society-Research Ireland University Research Fellowship.
\end{acknowledgements}

\bibliographystyle{aa} 
\bibliography{Lim_bib} 

\end{document}